\documentclass[a4paper]{proc}
\newcommand{\vek}[1]{\ensuremath{{\bf #1}}}
\title{Operational indistinguishably of varying speed of light theories}
\author{Nosratollah Jafari$^1$
  ~~and~~  Ahmad Shariati$^2$
\\[5pt] $^1$ \textit{Institute for Advanced Studies in Basic Sciences,}
\\[0pt] \textit{P.O. Box 159, Zanjan 45195, Iran}\\ \texttt{\normalsize njafary@iasbs.ac.ir}
\\[5pt] $^2$ \textit{Department of Physics, Alzahra University,}
\\[0pt] \textit{Tehran 19938-91167, Iran.}\\ \texttt{\normalsize shariati@mailaps.org}
} 

\begin{document}
\maketitle
\begin{abstract}
\vspace{0.5cm} The varying speed of light theories  have been
recently proposed to solve the standard model problems and
anomalies in the ultra high energy cosmic rays. These theories try
to formulate a new relativity with no assumptions about the
constancy of the light speed. In this regard, we study two
theories and want to show that these theories are not the new
theories of relativity, but only re-descriptions of Einstein's
special relativity.
\end{abstract}
\vspace{0.5cm}
\vspace{0.5cm}
\section{Introduction}
\vspace{0.5cm}
The varying speed of light theories has taken much
attraction recently \cite{Mag}. These theories have been proposed
to solve some problems in the standard model cosmology,
\textit{viz.}\  the horizon and the flatness problems, and the
anomalies in ultra high energy cosmic rays and TeV photons. The
idea is that, may be Einstein's special relativity needs some
modifications in very high energies or very large scales.

Einstein's special relativity is based on the following two principles:
\begin{enumerate}
\item The relativity principle: The physical laws are the same in
all inertial frames.
\item  The constancy of the speed of light: The speed of light is
independent of the speed of observer and source.
\end{enumerate}
These two principles, combined with some assumptions about the homogeneity of
space and time, and isotropy of space, lead to Lorentz invariance.

One may ask the following question: ``What if one relaxes
Einstein's second postulate?'' That is, what if one assumes that
the speed of light, being finite, need not to be constant.  There
are some arguments, showing that if one assumes that the speed of
light is finite, there is no way out of its being a universal
constant \cite[pp. 1--9]{SU}.  In other words, there is no way out
of the Lorentz group being the local symmetry group of our
spacetime.

Now, there are some efforts to construct theories, alternative to
the special relativity, in which the speed of light is not a
constant. These are called ``varying speed of light'' theories
(VLS). The Fock--Lorentz (FL) \cite{F} and Magueijo--Smolin (MS)
\cite{MS1} transformations are two important examples of such
theories. This article is a set of comments about these theories.

The main point is that these theories are basically re-description
of Einstein's special relativity in non-cartesian coordinates
coordinates. This result is in agreement with those found by
Ahluwalia-Khalilova \cite{AHL1,AHL2} and Grumiller $et. al$
\cite{Grum}. \vspace{0.5cm}
\section{Fock--Lorentz Transformations }
These transformations were obtained firstly by Fock \cite{F} with
no assumption about the constancy of light speed. In Fock's
argument the most general transformations between two systems that
are in uniform motion with respect to each other are
linear-fractional transformations. Recently, the same
transformations are re-derived by Stepanov \cite{S1,S2} and Manida
\cite{Man}.

The FL transformations between systems $S$ and $S'$ that are in
relative motion with constant speed $v$ (along the common
$x$-axis) are:
\begin{eqnarray}
\label{eq1}t'=\frac{\gamma(t-\frac{v}{c^{2}}x)}{1+\lambda v\gamma
x- \lambda
c^{2}(\gamma-1)t},   \\
x'=\frac{\gamma(x-vt)}{1+\lambda v\gamma x- \lambda
c^{2}(\gamma-1)t}, \\
y'=\frac{y}{1+\lambda v\gamma x- \lambda
c^{2}(\gamma-1)t},   \\
\label{eq4}z'=\frac{z}{1+\lambda v\gamma x- \lambda
c^{2}(\gamma-1)t},
\end{eqnarray}
where $\gamma=(1-\frac{v^{2}}{c^{2}})^{-\frac{1}{2}}$,
$c=299,792,458\,{\rm m}{\rm s}^{-1}$, and $\lambda$ is a quantity
with dimensions $L^{-2}T$, related to the Hubble's constant
through $H \propto \lambda c^2$.  If we assume $H \simeq 10^{2}
{\rm km}\, {\rm sec}^{-1} \, {\rm Mpc}^{-1}$, we have $\lambda
\sim 10^{-35} {\rm m}^{-2}\, {\rm s}$, which shows that for all
conceivable lengths and velocities, the denominators are very
close to $1$.  Therefore, these transformations are very close to
the usual Lorentz transformations.

The FL transformations form a group, which is isomorphic to the
Lorentz group.  In fact, the FL transformations are a
\textit{non-linear} representation of the Lorentz group.

 The most important difference between FL and usual
Lorentz transformations is the variability of light speed. The
speed that left invariant by the FL transformations instead of $c$
is
\begin{equation} \label{eq3}
C(t,\vek{x})=\left\vert\frac{\vek{c}+\lambda\, c^{2}\,\vek{x}}{1+\lambda\,
 c^{2}\, t}\right\vert
\end{equation}
here $\vek{c}=c\, \hat{n}$, where $\hat{n}$ is a unit vector, and
$c$ is the constant $299,792,458\,{\rm m}{\rm s}^{-1}$; and
$C(t,\vek{x})$ is a function of space and time. $C$ is a
decreasing function of time \cite{S1}.

It is easily seen from the FL transformations that these
transformations left invariant the expression
\begin{equation}
\frac{c^{2}\,t^{2}-\vek{x}\cdot\vek{x}}{(1+\lambda\, c^{2}\,t)^{2}}
\end{equation} that is
\begin{equation}
\frac{c^{2}t^{2}-\vek{x}\cdot\vek{x}}{(1+\lambda\,c^{2}\,t)^{2}}
=\frac{c^{2}\,t'^{2}-\vek{x}'\cdot\vek{x}'}{(1+\lambda\,
c^{2}t')^{2}}.
\end{equation}

Now, it is well known \cite{S1} that the FL spacetime is a Lorentzian
spacetime with the following metric:
\begin{eqnarray}
ds^{2} = \frac{1-\lambda^2 \, c^2 \, \vek{x}\cdot\vek{x}}{(1+\lambda \, c^{2}t)^{4}}
\, c^{2}\, dt^{2}  &+&
\frac{2\lambda \, c^{2}\,\vek{x}\cdot\, d\vek{x}\, dt}{(1+\lambda c^{2}t)^{3}}
\cr & - & \frac{d\vek{x}\cdot d\vek{x}}{(1+\lambda c^{2}t)^{2}}
\end{eqnarray}
Although this is a rather complicated metric, it can be shown that
its Riemann tensor vanishes identically. Therefore, this spacetime
is locally isometric to a part of the Minkowski spacetime. In
fact, in the new coordinates
\begin{eqnarray}
&& T=\frac{t}{1+\lambda c^{2}t},
\\ && \vek{X}=\frac{\vek{x}}{1+\lambda c^{2}t}
\end{eqnarray}
the metric reads
\begin{eqnarray}
ds^{2}=c^{2}dT^{2}-dX^{2}-dY^{2}-dZ^{2}.
\end{eqnarray}

\begin{figure}
\begin{center}
\begin{picture}(76,76)(0,0)
\multiput(0,0)(76,0){2}{\line(0,1){76}}
\multiput(0,0)(0,76){2}{\line(1,0){76}}
\includegraphics{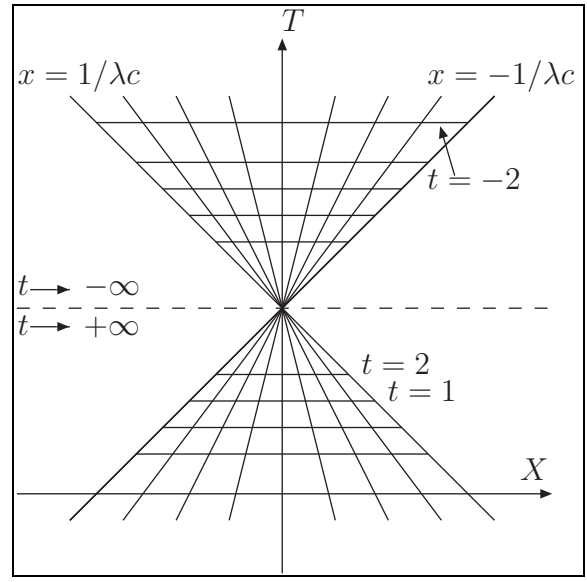}
\end{picture}
\end{center}

\caption{FL spacetime  diagram in terms of $(T,, X)$ coordinates.
FL coordinates $t$ and $x$ is drawn as horizontal and radial lines.}
\end{figure}

\par
 Let's investigate the meaning of this transformation.
Note that $T$ is the temporal coordinate, and $\vek{X} = (X, Y,
Z)$ are cartesian coordinates, in an inertial frame. Coordinates
$t$ and $\vek{x}:=(x, y, z)$ are being defined through these
relations. Inverting this transformation, we get $t=T/(1 -
\lambda\, c^2\, T)$, $\vek{x}=\vek{X}/(1 - \lambda\, c^2\, T)$. It
is obvious that these transformations are not well defined on the
hyperplane $T =1/\left(\lambda\, c^{2}\right)$. This hyperplane is
illustrated as a dashed line in fig 1. From $T =t/(1+\lambda\, c^2
\, t)$, or its equivalent $t=T/(1 - \lambda\, c^2\, T)$, we know
that the sections $T=$~constant, are the same as the sections
$t=$~constant; only the labelling is different. Note that the
section $T =1/\left(\lambda\, c^{2}\right)$, is a hyperplane of
discontinuity in $t$. Now from $\vek{x}=\vek{X}/(1 - \lambda\,
c^2\, T)$, we see that the lines $\vek{x}=(x_0, y_0, z_0)$ are the
lines $\vek{X} = \vek{x} - \lambda \, c^2 \, \vek{x}\, T$. This
means that, as far as $\lambda\, c^2\, \vert\vek{x}\vert < c$, one
can interpret the line $\vek{x}=(x_0, y_0, z_0)$ as the worldline
of a massive particle, or an observer. In other words, the set of
lines $-\frac{1}{\lambda\, c} < \vert\vek{x}\vert <
\frac{1}{\lambda\, c}$, which lie inside the lightcone with vertex
$(T=1/\left(\lambda\, c^2\right), X=0, Y=0, Z=0)$, can be
interpreted as the worldlines of point observers. Therefore,
inside the lightcone indicated in Fig. 1, one can interpret
$\vek{x}$ and $t$ as spacial and temporal coordinates.


In terms of these new \emph{time and space coordinates}, the FL transformations
(\ref{eq1}-\ref{eq4}) reads
\begin{eqnarray}
&& T'=\gamma  (T-\frac{v}{c^{2}}X), \\ && X'=\gamma(X-v T),
\\ && Y'=Y, \\ && Z'=Z
\end{eqnarray}
This is the usual Lorentz transformations for $(T,X,Y,Z)$
coordinates. Thus, the FL transformations is solely a
re-description of Einstein's special relativity in the unusual
coordinates. Although this is quite well known, we think people do
not pay enough attention to it.



It is useful to mention that the FL spacetime is the same as Milne
spacetime. The Milne spacetime metric is
$ds^{2}=d\tau^{2}-\tau^{2}d\beta^{2}$ and its spacetime diagram is
shown in Fig.2.
\begin{figure}
\begin{center}
\begin{picture}(76,76)(0,0)
\multiput(0,0)(76,0){2}{\line(0,1){76}}
\multiput(0,0)(0,76){2}{\line(1,0){76}}
\includegraphics{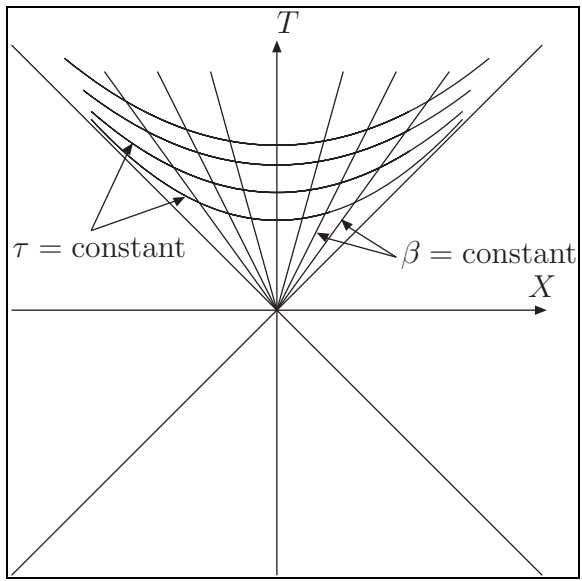}
\end{picture}
\end{center}

\caption{ The Milne spacetime  diagram in terms of $(T ,X)$ coordinates.}
\end{figure}
In Fig.2 the $\tau $-constant hyperbolae and the $\beta$-constant
lines specify the Milne coordinates. The relation between the
Milne coordinates $(\tau,\beta)$ and the Minkowskian coordinates
$(T,X)$ is
\begin{eqnarray}
&& \tau=\sqrt{T^{2}-X^{2}},\\ && \tanh\beta=\frac{X}{T}.
\end{eqnarray}
By means of these relations one can obtain the relation between
the FL coordinates $(t,x)$ and the Milne coordinates
$(\tau,\beta)$.
\begin{eqnarray} \label{eq12a}
&& \tau=\frac{\sqrt{t^{2}-x^{2}}}{1+\lambda c^{2}t},
\\ \label{eq12b}
 && \tanh\beta=\frac{x}{t}.
\end{eqnarray}

All phenomena that are mentioned about the FL spacetime can also
be seen in the Milne spacetime. Thus it seems that light speed is
variable in this spacetime, too. Let us define the light speed as
coordinate speed. From
\begin{eqnarray} ds^{2}=d\tau^{2}-\tau^{2}d\beta^{2}=0,
\end{eqnarray}
we get
\begin{equation}
c_{light}:=\frac{d\beta}{d\tau}=\frac{1}{\tau}.
\end{equation}
With this definition, the light speed is large at initial moments
and decrease with Milne time $\tau$. Therefore, the light speed
that is variable in Stepanov and Manida viewpoint is the
coordinate speed.

It is interesting to consider some particles in the origin of the
Milne spacetime at $\tau =0$. These particles move away in the
direction of the $\beta$-coordinates with passing time. This
picture is the Milne enlarging universe picture and the rate of
this enlargement depends on the $\lambda$ parameter, which is the
Stepanov's \cite{S2} result about the Hubble's constant $H \propto
\lambda
 c^{2}$.

\section{Magueijo-Smolin Transformations}
Magueijo and Smolin generalize Fock's idea to the momentum space
and found transformations that left invariant the Planck energy
\cite{MS1,MS2}. Note that by the usual linear action of the
Lorentz boosts on the four-momenta of particles, the Planck energy
is not invariant. However, one can define a non-linear action of
the Lorentz boosts on the four-momenta of particles, such that a
specific energy scale, to be called the Planck scale, is
invariant. When systems $S$ and $S'$ are in relative motion with
constant speed $v$ (along the common  $x$-axis), these non-linear
transformations read
\begin{eqnarray}
\label{eq2a}
p_{0}'=\frac{\gamma(p_{0}-\frac{v}{c}p_{x})}{1+\ell(\gamma-1)p_{0}-\ell\gamma \frac{v}{c}p_{x}},
\\ \label{eq2b}
p_{x}'=\frac{\gamma(p_{x}-\frac{v}{c}p_{0})}{1+\ell(\gamma-1)p_{0}-\ell\gamma \frac{v}{c}p_{x}},
\\ \label{eq2c}
p_{y}'=\frac{p_{y}}{1+\ell(\gamma-1)p_{0}-\ell\gamma \frac{v}{c}p_{x}},
\\ \label{eq2d}
p_{z}'=\frac{p_{z}}{1+\ell(\gamma-1)p_{0}-\ell\gamma \frac{v}{c}p_{x}} ,
\end{eqnarray}
where $\ell$ is a quantity with dimensions $M^{-1} L^{-1} T$,
related to the Planck length $l_p$ through $\ell\propto
l_p/\hbar$. (From now on we put $\hbar=1$, so that $\ell$ is the
Planck length.)

These transformations are also, a non-linear representation of the
Lorentz group in the momentum space, and depend on the ratio of
the particle energy to Planck energy. If the particle energy is
very small compared to the Planck energy, then these
transformations become the usual Lorentz transformations. But
looking at these transformations, we can see that changing
4-momentum $p_{\mu}$ to
\begin{eqnarray}\label{eq26}
P_{\mu}=\frac{p_{\mu}}{1-\ell p_{0}},
\end{eqnarray}
 then MS transformations become
\begin{eqnarray}
&& P'_{0}=\gamma (P_{0}-\frac{v}{c}P_{x}),
\\ && P'_{x}=\gamma (P_{x}-\frac{v}{c}P_{0}),
\\ && P'_{y}=P_{y},
\\ && P'_{z}=P_{z}
\end{eqnarray}
These are, the same usual Lorentz transformations for momentum
space. Therefore, the MS transformations, like the FL ones, are
only re-description of the usual Lorentz transformations in the
unusual coordinates. Also, it is easily seen that MS  momentum
space is the interior of the light-cone in the Minkowski momentum
space. The MS momentum space diagram is a figure like Fig.1.
 Also, as in the FL case all MS  results can be
 obtained from Einstein's special relativity by using the
 coordinate transformations (\ref{eq26}). For example, $ \|p\,\|^{2} $
 invariant in the MS  formalism can  be viewed as
 \vspace{0.4cm}
\begin{eqnarray}
\|p\,\|^{2}=\frac{\eta_{\mu\nu}p_{\mu}p_{\nu}}{(1-\ell
p_{0})^{2}}=\eta_{\mu\nu}P_{\mu}P_{\nu}= \|P\,\|^{2},
\end{eqnarray}
where $\eta_{\mu\nu}=\rm diag(-1,1,1,1)$ is the Minkowski metric.

After giving the momentum transformations (\ref{eq2a}--\ref{eq2d})
by Magueijo \cite{MS1}, some people like Mignemi \cite{Mig} and
Magueijo himself \cite{Kim} tried to find corresponding
transformations in spacetime. Although, they approached the
problem in different ways, they obtained the same result. Their
transformations between two observers that are in relative motion
with constant speed $v$ in two dimensions are
\begin{eqnarray}
t'=\Delta(t\cosh\xi-x\sinh\xi), \\
x'=\Delta(-t\sinh\xi+x\cosh\xi),
\end{eqnarray}
where
\begin{eqnarray}
\Delta=1+p_{0}\ell (\cosh\xi-1)+p_{1}\ell \sinh\xi,
\end{eqnarray}
$\xi$ is the rapidity parameter and $\ell $ is the Planck length.
The problem with these transformations is their energy and
momentum dependence: Spacetime is a set of events to which we
usually do not assign energy and momentum. Events occur in
definite positions and times and differ from particles that have
definite energies and momenta. Thus, the energy and momentum
dependency of the spacetime transformations picture is hard to be
meaningful.

However, the reverse procedure has no difficulty. For example, we
can look for the momentum transformations corresponding to the FL
spacetime transformations. In this case we have for two dimensions
\begin{eqnarray}
p_{0}'=&{}& \tilde{\Delta}\;\left[\cosh\xi+\lambda t\,(\cosh\xi-1)\right]p_{0}
\cr & + & \tilde{\Delta}\;\left[\sinh\xi+\lambda x\,(\cosh\xi -1)\right] p_{1}, \\
\cr p_{1}'=&{}& \tilde{\Delta}\;\left[\sinh\xi+\lambda t\, \sinh\xi\right]p_{0}
\cr &+& \tilde{\Delta} \;\left[\cosh\xi+ \lambda x\sinh\xi)\right] p_{1},
\end{eqnarray}
where
\begin{eqnarray}
\tilde{\Delta}=1+\lambda x\sinh\xi-\lambda(\cosh\xi-1)t,
\end{eqnarray}
and $\lambda$ is a small quantity related to Hubble's constant.
 In these transformations, momenta depend on the position and time.
 This is not absurd, because a classical particle, at each
 instant, is at a definite position.

\section{Conclusion}
It is outlined that the MS and FL transformations are only
re-descriptions of Einstein's special relativity in one portion of
the Minkowski spacetime. In other words, Einstein's special
relativity still remains valid, and these transformations are only
descriptions of the special relativity in the language of
non-cartesian coordinates that lead to the apparent variability of
the speed of light, and apparent violation of the Lorentz
invariance.

\section{Acknowledgements}
We should thanks our colleagues M. Khorrami and A.H. Fatollahi at
IASBS for their kind discussions. We are also grateful to D.V.
Ahluwalia-Khalilova, A.F. Ranada, D.V. Vassilevich, R. Lehnert and
D. Grumiller for helping us put our work in perspective.

\end{document}